\documentclass[global]{svjour}
\usepackage{graphicx}
\usepackage{amssymb, amsmath}
\usepackage{lscape}
\usepackage{color}
\journalname{\emph{Astrophysics}}
\begin{document}
\title{CLOSE NEIGHBORS OF MARKARIAN GALAXIES}
\subtitle{II. STATISTICS AND DISCUSSIONS}
\titlerunning{CLOSE NEIGHBORS OF MRK GALAXIES. II.}
\author{T.~A.~Nazaryan\inst{1}
\and A.~R.~Petrosian\inst{1}
\and A.~A.~Hakobyan\inst{1}
\and B.~J.~McLean\inst{2}
\and and D.~Kunth\inst{3}
}                     
%
%
\institute{Byurakan Astrophysical Observatory, Armenia; \email{nazaryan@bao.sci.am}
\and Space Telescope Science Institute, USA
\and Institut d'Astrophysique de Paris, France}
%
%
\maketitle
\begin{abstract}
\begin{center}
\parbox{0.9\hsize}{\emph{According to the database from the first paper, we select 180 pairs with $\emph{d}V < 800~ km~ s^{-1}$
and $D \emph{p} < 60~ kpc$ containing Markarian (MRK) galaxies. We study the dependence of galaxies’
integral parameters, star-formation (SF) and active galactic nuclei (AGN) properties on
kinematics of pairs, their structure and large-scale environments. Following main results were
obtained: projected radial separation $D \emph{p}$ between galaxies correlates with the perturbation level
P of the pairs. Both parameters do not correlate with line-of-sight velocity difference $\emph{d}V$ of
galaxies. $D\emph{p}$ and P are better measures of interaction strength than $\emph{d}V$. The latter correlates with
the density of large-scale environment and with the morphologies of galaxies. Both galaxies in a
pair are of the same nature, the only difference is that MRK galaxies are usually brighter than
their neighbors in average by 0.9~mag. Specific star formation rates (SSFR) of galaxies in pairs
with smaller $D\emph{p}$ or $\emph{d}V$ is in average 0.5~dex higher than that of galaxies in pairs with larger $D\emph{p}$ or
$\emph{d}V$. Closeness of a neighbor with the same and later morphological type increases the SSFR,
while earlier-type neighbors do not increase SSFR. Major interactions/mergers trigger SF and
AGN more effectively than minor ones. The fraction of AGNs is higher in more perturbed pairs
and pairs with smaller $D\emph{p}$. AGNs typically are in stronger interacting systems than star-forming
and passive galaxies. There are correlations of both SSFRs and spectral properties of nuclei
between pair members.
}}
\end{center}
\end{abstract}
\begin{center}
\parbox{0.9\hsize}{\keywords{\emph{galaxies: general -- galaxies: interactions -- galaxies: starburst -- galaxies:
active -- galaxies: peculiar}}}
\end{center}

\section{Introduction}
\label{intro}

Close interactions/mergers of galaxies are considered as important
processes influencing morphological, stellar and chemical evolution of galaxies.
Numerous observational results show that interactions/mergers trigger SF in galaxies.
The pioneering work \cite{larson78} showed that peculiar galaxies have
wider spread on color-color diagram and, generally,
are bluer than normal galaxies.
The authors suggested that sharp bursts of SF,
with their timescales consistent with interactions,
can explain peculiar colors of these galaxies.
Later, many others showed that closer pairs of galaxies have
enhanced star formation rates (SFRs) measured by
emission lines, e.g. \cite{barton00,lambas03,li08,ellison08},
optical colors, e.g. \cite{darg10,patton11},
infrared (IR) emission, e.g. \cite{hwang10,ellison13},
supernovae distribution, e.g. \cite{petrosian95,nazaryan13}.
The main physical processes responsible for the enhanced SF are
gas inflow toward nuclear regions of galaxies due to global torques and,
probably, gas fragmentation into massive and dense clouds and rapid SF therein, e.g. \cite{mihos96,hopkins13}.
The triggering mechanism of AGN is often considered to be the same as that
of the enhanced nuclear SF \cite{ellison08,woods07,wild10,ellison11a,liu12}.

In spite of that, various aspects and many factors can affect on
frequency and efficiency of enhanced SF triggering by galaxy interaction and merging.
The role of large-scale environment is still debated, especially,
taking into account morphology-density relation \cite{dressler80}.
In this respect, in \cite{alonso06,ellison10} it was observed that galaxy interactions are
more effective in triggering SF in low- and moderate-density environments.
In addition, in \cite{hwang10,park09} it was found that
late-type neighbors enhance SF of galaxies while early-type neighbors reduce it,
and it was showed that the role of the large-scale density in determining galaxy properties
is minimal once luminosity and morphology are fixed.
On the other hand, effect of large-scale environment
was considered small, but non-zero in \cite{scudder12a}.
It is mostly assumed that major mergers are more effective in triggering starbursts (also AGNs),
than minor ones \cite{li08,ellison08,woods07,ellison11a,alonso07,cox08,scudder12b,lambas12}.
At the same time, minor mergers occur more frequently, and partially can explain
triggered SF in early-type galaxies, e.g. \cite{hakobyan08,kaviraj09}.

Finally, the general role of interactions and mergers
in triggering SF of galaxies is still not clear.
The facts, that not all galaxies with high SFR are interacting ones,
as well as that not all interacting galaxies have high SFR,
support the hypothesis that internal properties of galaxies are also
an important factor determining enhanced SF \cite{lambas03,alonso06},
especially at higher redshifts \cite{dimatteo08,xu12}.
Bars, transferring gas to nuclei, can be an alternative mechanism of induced nuclear SF \cite{ellison11b,adibekyan09},
although bars themselves are disputably considered to be interaction-induced, e.g. \cite{casteels13}.

The large variety of parameters is one of the main difficulties in studying
influence of interactions and mergers on SF and AGN properties of galaxies.
The choice of galaxies as interacting also contains some ambiguities,
because interacting pairs can be selected by different criteria,
such as according to the difference of line-of-sight (LoS) velocity,
projected distance between pair members, or the degree of morphological disturbances
(assessed both visually or automatically, e.g. by asymmetry).
Although these parameters are correlated with each other at first approximation, see e.g. \cite{lambas12,patton00},
they can bias the pair statistics in different ways
because of correlation with large-scale environment, e.g. \cite{ellison10},
possible effects on morphological classification, e.g. \cite{casteels13},
or different timescales and sequences of SF, AGN and disturbed morphology \cite{ellison08,darg10,hopkins13,hopkins12}.
The sizes of samples also vary greatly in different studies,
from several hundreds to a hundred thousand \cite{li08},
bringing additional difficulties for making satisfactory conclusions.
Additional, scrupulously chosen samples can provide further results to reveal the details.

The aim of this study is to investigate the connections between gravitational interaction with a close neighbor
and nuclear activity and/or enhanced SF in galaxy pairs.
We will study these phenomena through examining dependence of integral parameters,
SFRs and nuclear properties of galaxies in pairs on kinematical properties of these systems,
as well as their largescale environments.
Also, we will investigate correlations between properties of paired galaxies.

In our first paper of a series \cite{nazaryan12}, we have reported the creation of
the database of close neighbors of MRK galaxies, which contains extensive
new measurements of their optical parameters, collected near-IR data, and pair properties.
This is a second paper of a series and the outline is as follows:
Sect.~\ref{sample} presents the sample of close pairs of galaxies and selected parameters.
Section~\ref{statistics} presents the statistical study of the sample and discusses its results.
Section~\ref{summary} is the summary of this study.
Throughout this paper, we adopt the Hubble constant
$H_{\rm{0}} = 73~\rm{km~ s^{-1}~Mpc^{-1}}$.

\section{The sample}
\label{sample}

There are two possible approaches for pair selection with the purposes of
studying SF and AGN properties of pair members.
First, one can select all pairs from a catalog of galaxies
and study SF and AGN properties in these pairs.
Second, one can choose only galaxies with desirable properties,
search for their neighbors and study properties of these pairs.
The first approach is more commonly used, e.g. in \cite{ellison11a},
because it provides selection of pairs with all possible variety of parameters
and is limited only by original catalogs of galaxies drawn from.
The second approach (e.g. used in \cite{liu12}) gives an opportunity to select
a well-chosen sample of pairs, which generally will be smaller,
but the selection effects should be constrained in a comprehensible manner.
We adopt the second approach in our pair selection.
It gives us an opportunity to select well-chosen sample containing
enough number of galaxies with different activity levels,
but small enough to study it thoroughly via visual classification
and manual checking of automatically measured parameters.

The starting point to create our sample of pairs is the catalog of MRK galaxies.
The original catalog \cite{markarian89} features 1545 bright galaxies
mostly having starburst properties and/or AGNs, which are well-studied objects.
In \cite{petrosian07}, homogeneously measured parameters of MRK galaxies,
such as magnitudes, sizes, positions, redshifts, and morphologies, are presented.

Results of a close neighbors search for MRK galaxies within position-redshift space
using the NASA/IPAC Extragalactic Database and Sloan Digital Sky Survey (SDSS) Data Release (DR) 8
(for current study we added to this list some new objects found only in the DR9 \cite{ahn12})
are published in \cite{nazaryan12}.
In \cite{nazaryan12}, three criteria were used to select the sample of close neighbors of MRK galaxies.
(1)~Redshift of MRK galaxy should be more than 0.005.
(2)~Difference of LoS velocities of MRK galaxy and its neighbor should be less than $\rm{800~km~s^{-1}}$.
(3)~Projected distance between MRK galaxies and their neighbors should be less than $\rm{60~kpc}$ (close systems).
According to these criteria, 633 galaxies in close systems containing 274 MRK galaxies were found.
For the current study, only pairs of galaxies were selected from above mentioned 633 galaxies.
The total number of pairs containing at least one MRK galaxy is 217.
The percentage of pairs in our sample located within the SDSS coverage is 83~\%.
The median distance of selected pairs is 96~Mpc.

\begin{figure}[t]
\centering
\includegraphics[width=0.5\hsize]{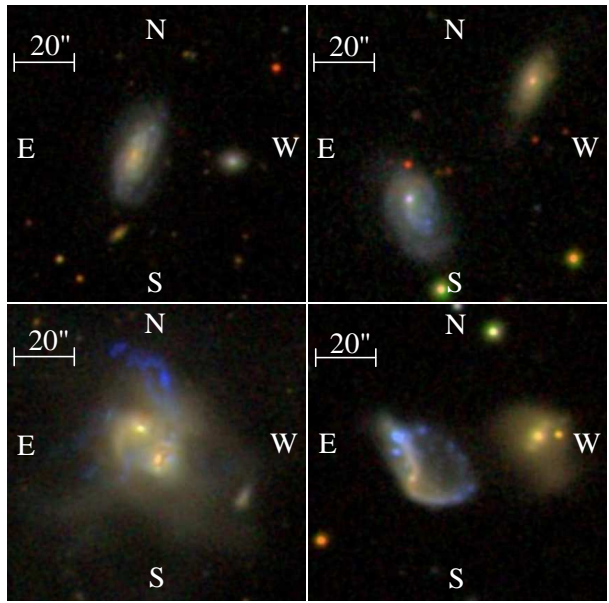}\\
 \parbox{0.5\hsize}{\caption{Examples of pairs with perturbation levels
 $P=0$ (top left), $P=1$ (top right), $P=2$ (bottom right),
 and $P=3$ (bottom left).}\label{perturbations}}
\end{figure}

The morphological classification of MRK galaxies and their neighbors was done
in \cite{petrosian07} and \cite{nazaryan12} respectively,
using only Digitized Sky Survey-II (DSS-II) blue and red images for MRK
and DSS-II blue and red images as well SDSS color images for neighbors.
For this study, we check morphological classes of MRK galaxies
by using also SDSS color images for homogeneity.
New and revised morphological classes were suggested for the 16~\% of MRK galaxies,
morphological classes were specified for other 17~\% of galaxies.
The 19~\% of sample galaxies are of early-types (earlier than S0/a),
46~\% of early spirals (Sa-Sbc), and 36~\% of late spirals and irregulars (later than Sc).

We classified sample pairs in terms of morphological perturbations, e.g. \cite{liu12,lambas12},
by 4 levels: $P = 0$: unperturbed pairs, $P = 1$: slightly perturbed, $P = 2$: highly perturbed, $P = 3$: mergers.
Unperturbed pairs are defined as having both components with no visible morphological perturbation.
Slightly perturbed pairs are pairs where the most perturbed component
has visible morphological perturbations, but without long tidal arms, bridges,
or violation of spiral patterns or brightness profile.
In highly perturbed pairs the most perturbed component has
significant morphological perturbations, such as long tidal arms, bridges,
or violation of spiral patterns or brightness profile.
Mergers are pairs with obvious merging processes.
The typical examples of pairs of each category are shown in Fig.~\ref{perturbations}.
The 45~\% of the pairs have $P = 0$, 23~\% $P = 1$, 22~\% $P = 2$, and 10~\% $P = 3$.
Blind reclassification of a sample shows that the selection criteria for perturbation
are quite reliable and objective, and errors of this classification are
less than errors of morphological \emph{t}-types classification.
Typically less than 20~\% of pairs change their $P$ level by one unit when reclassified.

In \cite{nazaryan12,petrosian07}, isophotal magnitudes of the galaxies from
DSS-II blue (\emph{J}), red (\emph{F}), and near-IR (\emph{I}) images were measured in an homogeneous way.
In this study we used SDSS \emph{cmodel} magnitudes for luminosities and \emph{model} magnitudes for colors,
which were controlled by DSS magnitudes (outliers were removed).

We describe the large-scale environments of each pair by average large-scale density $\Sigma$
calculated as a surface density of galaxies from SDSS within 1~Mpc projected circles
and with LoS velocity differences less than $\rm{1500~km~s^{-1}}$
(typical $\Sigma$ of galaxy groups is less than $\rm{500~km~s^{-1}}$ \cite{yang07}).
For that purpose, we counted all galaxies in absolute magnitude limited sample
$M \leq -18.74$ in $r$-band corresponding to SDSS completeness magnitude $r_{\rm{Petro}} < 17.77$
\cite{strauss02} at 200~Mpc distance (95~\% of pairs are closer).
So 95~\% of pairs do not lose galaxies in their 1~Mpc environments because of magnitude limit.
In higher density environments there is a systematic undercount of neighbors due to fiber collisions \cite{strauss02}.
We complete number of neighbor galaxies proportionally to not-covered area for
17 pairs having 1~Mpc circles partially located outside of SDSS coverage.
For our further statistics we divide all pairs according to $\Sigma$ into three categories:
low-, medium-, and high-density environments with $\Sigma \leq 2$, $2 < \Sigma \leq 5$, and $\Sigma > 5$ accordingly.

For statistics we also included some spectral parameters of galaxies processed by MPA-JHU pipeline,
which fits galaxy templates and spectral synthesis models to the spectra \cite{kauffmann03,brinchmann04}.
These parameters are SFR and SSFR for whole galaxy, and nuclear emission-line classification.
We visually inspected each galaxy in our sample and filtered
only those having nucleus located within SDSS spectral fiber.
Mean values of $log({\rm SFR})$ and $log({\rm SSFR})$ are $-0.1 \pm 0.8$ and $-9.9 \pm 0.9$
for MRK and $-0.7 \pm 0.9$ and $-10.1 \pm 0.9$ for neighbors.

We divide galaxies of our sample into four groups basing on nuclear
BPT \cite{baldwin81} classification of their SDSS spectra.
These classes are ${\rm BPT = PS}$ for passive nuclei (which are not included in BPT classification);
${\rm BPT = SF}$ for star-forming nuclei; ${\rm BPT = C}$ for composite nuclei, ${\rm BPT = AGN}$ for Seyfert galaxies, and ${\rm BPT = L}$ for LINERs.
The $9 \pm 2~\%$ of galaxies out of 180 with available spectra have passive,
$66 \pm 6~\%$ have star-forming, $14 \pm 3~\%$ have composite, $2 \pm 1~\%$ have AGN, and $8 \pm 1~\%$ have LINER nuclei.
Surprisingly, we found 7 cases when the nucleus of MRK galaxy has spectral classification ``passive''.
These galaxies are: MRKs 422, 562, 654, 842, 902, 1276, and 1349.
We inspected their SDSS spectra and found that only MRK~654 has emission lines
and can have typical spectrum of starburst galaxy.
All the other cases are early-type galaxies with neither excess in blue band nor any strong emission lines.

We study some possible selection effects of the sample that could bias further statistics.
The dependence of absolute magnitude on redshift (\emph{Malmquist} bias)
is quite strong in the sample: $M_{\rm r} = -18.45 - 0.0137~d$;
(correlation coefficient $ r = -0.50$), where $d$ is the distance from us in Mpc.
As a result, morphologies of galaxies are also biased by distance:
$d = 111 \pm 67~ \rm{Mpc}$ for early-types, $d = 119 \pm 59~ \rm{Mpc}$ for early spirals,
and $d = 82 \pm 42~\rm{Mpc}$ for late spirals and irregulars.
Other morphological features, i.e., perturbation levels of pairs and bar detection, are not biased by distance.
Kinematical parameters of pairs, i.e. ${\rm d}V$ and $D\rm{p}$ are also not biased by distance.
Because of sample selection criteria, we have deficit of passive-passive pairs
and, to a lesser extent, active-passive pairs.

\section{Statistics and discussions}
\label{statistics}

\subsection{Multivariate factor analysis (MFA)}

The statistical research was conducted in two steps.
First, we applied an exploratory MFA to look for correlations between all
parameters describing MRK galaxies, their neighbors (\emph{t}-type, bar, abs. mag, $B - R$, SSFR, and BPT type)
and pair properties (${\rm d}V$, $D\rm{p}$, $P$, and $\Sigma$).
This statistical method is similar to the more commonly used principal component analysis.
The MFA describes the interdependence and grouping patterns of variables in terms of factors.
Factor loadings are measures of involvement of variables in factor patterns
and can be interpreted like correlation coefficients.
The square of the loading is the variation that a variable has in common with the factor pattern.
The percent of total variance carried by a factor is the mean of squared loadings for a factor.
In order to simplify the interpretation of the results, we only present
the rotated varimax normalized orthogonal values for the three most significant factors,
with highlighted values above 0.4 correlation threshold.
Table~\ref{MFAtable} shows the factor loadings, i.e., the correlation coefficients
between the initial variables and the factors for the $N = 59$ subsample
with known values of all initial variables of galaxies.

\begin{table}[t]
\begin{center}
\parbox{0.4\hsize}{\caption{Varimax rotated normalized orthogonal factor loadings.}
\label{MFAtable}}\\
\begin{tabular}{lrrr}
\hline
\hline
Variable & \multicolumn{1}{c}{$F_1$} & \multicolumn{1}{c}{$F_2$} & \multicolumn{1}{c}{$F_3$}\\
\hline
${\rm d}V$ & \bf--0.51&0.06& --0.09\\
$D\rm{p}$&--0.11& \bf0.48& --0.09\\
\emph{P}& --0.11& \bf--0.75& 0.31 \\
$\Sigma$& \bf--0.42&0.39& 0.15\\
\emph{t}-type MRK&\bf 0.80& 0.04 &0.21 \\
Bar MRK&0.16 & 0.30 & 0.36\\
\emph{M} MRK&\bf 0.70 &\bf 0.43& 0.12 \\
$B-R$ MRK &\bf --0.53& 0.12& --0.16\\
SSFR MRK&\bf 0.71& --0.02& 0.13\\
BPT MRK & 0.07&\bf --0.54& --0.23\\
\emph{t}-type Neig.&\bf0.77& --0.05 &--0.27 \\
Bar Neig.&--0.02 & --0.05 & \bf 0.72\\
\emph{M} Neig.& \bf 0.61&\bf 0.42& --0.38 \\
$B-R$ Neig. &\bf --0.49& 0.29& \bf 0.54\\
SSFR Neig.&\bf 0.77& --0.14& --0.30\\
BPT Neig.  & 0.12&--0.11& \bf--0.47\\
\hline
Accum. variance& 26~\% & 38~\% & 49~\%\\
\hline
\end{tabular}
\end{center}
\end{table}

Factor $F_{\rm 1}$ is the combination of LoS velocity difference ${\rm d}V$,
density of environment $\Sigma$ and \emph{t}-types, abs. mag, colors, SSFRs of MRK and neighbor galaxies.
Pairs with smaller ${\rm d}V$ and in less denser environments
have preferably fainter and bluer galaxies of later morphological types
and with higher SSFRs.
Factor $F_{\rm 2}$ is the combination of pair perturbation levels $P$, $D\rm{p}$ separation,
abs. mag of MRK and neighbor galaxies, BPT classification of MRK galaxy.
MRK galaxies with active nuclei are located in closer and more perturbed pairs,
they and their neighbors are relatively luminous galaxies.
Factor $F_{\rm 3}$ connects bar existence, color and nuclear activity of neighbor galaxies.
Redder neighbor galaxies have larger fraction of bars and active nuclei.
These results are expected, they show common trends connecting properties of galaxies
and their environments \cite{ellison11a,dressler80,ellison10}.

\subsection{Sample properties}

\begin{figure}[ht]
\centering
\includegraphics[width=0.6\hsize]{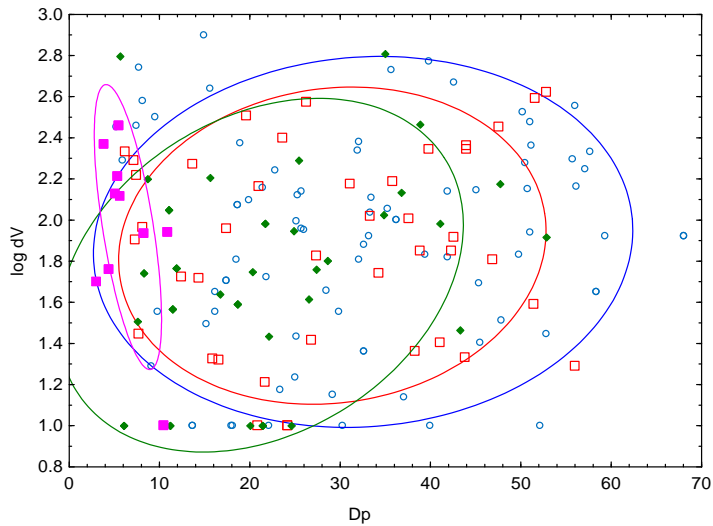}\\
\parbox{0.6\hsize}{\caption{$D\rm{p}$ vs. $log({\rm d}V)$ for all pairs with
different levels of perturbation $P$. $P=0$ are
marked by blue small blank circles, $P=1$ by
red large blank squares, $P=2$ by green small filled
diamonds, $P=3$ by purple large filled squares.
Ellipses correspond to 95~\% of points of
each distribution.}\label{dVDpP}}
\end{figure}

An important goal of this study is to examine the dependence of
SSFR and BPT types (target parameters) on ${\rm d}V$, $D\rm{p}$ and $P$ (primary parameters)
taking into account the impact of secondary parameters
such as morphologies, large-scale environments, luminosity ratio of pair components.
In this section we mention some other important relations
(between primary and secondary parameters)
that are essential to consider when discussing the sample properties.
The dependence of visually detected perturbation level $P$ on ${\rm d}V$ and $D\rm{p}$ is worthy to mention.
Figure~\ref{dVDpP} shows the distribution of pairs by their ${\rm d}V$ and $D\rm{p}$,
pairs with different perturbation levels $P$ are marked.
It is obvious and shown by MFA too, that $P$ correlates only with $D\rm{p}$,
while ${\rm d}V$ and $D\rm{p}$ do not correlate with each other.
The closer pairs are more disturbed.
This is the result of different nature of $D\rm{p}$ and ${\rm d}V$.
Pairs with larger ${\rm d}V$ correspond to environments with higher densities $\Sigma$.
Therefore while $D\rm{p}$ is a measure of interaction strength,
the variation of ${\rm d}V$ mainly reflects change of large-scale environments.

The fraction of barred galaxies in the sample depends neither on morphological type
(after removing elliptical and irregulars) nor on $D\rm{p}$.
On the other hand, there is a strong decrease
of number of barred galaxies from $\rm{47 \pm 10~\%}$ for small ${\rm d}V$ ($\rm{10 - 20~km~s^{-1}}$) pairs
to $\rm{14 \pm 3~\%}$ for large ${\rm d}V$ ($\rm{ > 100~km~s^{-1}}$) pairs.
The SSFRs of barred galaxies do not differ from those of unbarred ones significantly.

\subsection{MRK galaxies vs. neighbors}

We compared properties of neighbors with those of MRK galaxies.
Mean absolute \emph{cmodel r}~mag of neighbors is $-19.4 \pm 1.8$ compared to $-20.3 \pm 1.2$ for MRK galaxies.
Median morphological type of neighbor galaxy is Sbc compared to Sb for MRK galaxy.
A Kolmogorov-Smirnov (KS) test of morphologies gives $p = 0.09$ that
MRK and neighbors are drawn from the same sample.
Spearman's rank ($R = 0.29$) test shows that the morphologies of MRK and neighbors correlate significantly ($p = 0.00005$).
Number of barred galaxies in neighbors sample is $23 \pm 4~\%$ compared to $22 \pm 3~\%$ for that in MRK galaxies.
Comparison of SSFRs shows that SSFR of neighbors is not less than that of MRK galaxies.
The distribution of neighbors by BPT types is consistent with the distribution of MRK galaxies.
A KS test shows that the distribution of neighbors by colors is statistically the same as that of MRK galaxies.
Therefore neighbor galaxies are of the same nature as MRK galaxies.
We consider this fact as a result of existence of correlation between properties of galaxies in pairs.
Because of magnitude limitation of Markarian survey \cite{markarian89},
MRK galaxies are usually the brightest members of pairs
and are brighter in average by 0.9~mag.

\subsection{Dependence of SSFR on the parameters of interaction}

\begin{figure}[t]
\centering
\includegraphics[width=0.6\hsize]{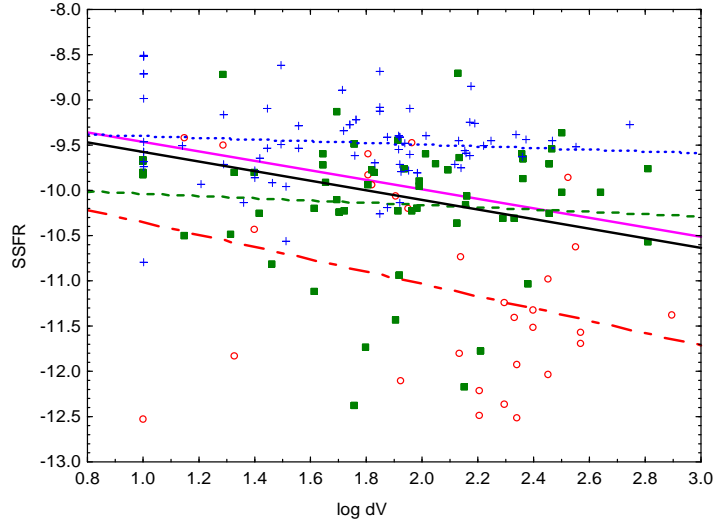}\\
\parbox{0.6\hsize}{\caption{SSFR vs. $log({\rm d}V)$ for subsamples
of early-types (red blank circles, dashed-dotted
line), early spirals (green filled squares,
dashed line), and late spirals and
irregulars (blue crosses, dotted line). Two
best-fit lines for all galaxies (black bottom solid
line) and AGN--removed sample (purple upper
solid line) are drawn.}\label{SSFRvsdVmorph}}
\end{figure}

The main parameters describing interactions are ${\rm d}V$, $D\rm{p}$, and $P$.
Figure~\ref{SSFRvsdVmorph} shows the dependence of SSFR on ${\rm d}V$ of a pair.
We remove galaxies with AGNs, then categorize the rest by morphologies to make adequate comparison.
Figure~\ref{SSFRvsdVmorph} shows that, without considering morphologies,
we see 0.7~dex increase of SSFRs from larger ${\rm d}V$ to smaller ones.
However the variance of SSFRs because of morphologies is much larger (more than 2.5~dex).
In \cite{ellison10}, it was shown that ${\rm d}V$ is biased by large-scale environment:
pairs in denser environments have larger ${\rm d}V$.
Because of morphology--density relation, ${\rm d}V$ is also biased by morphologies:
early-type galaxies have ${\rm d}V\sim250~\rm{km~ s^{-1}}$
while irregulars have ${\rm d}V\sim70~\rm{km~ s^{-1}}$,
so most of the SSFR vs. ${\rm d}V$ dependence is because of morphology--SSFR dependence
and does not reflect pure interaction.
Thus, we conclude that it is essential to take into account morphologies of galaxies
when discussing their SSFRs and interactions to obtain unbiased results.
Figure~\ref{SSFRvsdVmorph} shows that grouping by morphologies
weakens SSFR vs. ${\rm d}V$ relation, but there still remains some variance,
which is maximal for early spirals ($0.4 - 0.5$~dex).
This result is in agreement with modeling, e.g. \cite{dimatteo08}
showing that strong starbursts during interactions are rare
and that typical enhancement of SF is less than 5 times.

\begin{figure}[t]
\centering
\includegraphics[width=0.6\hsize]{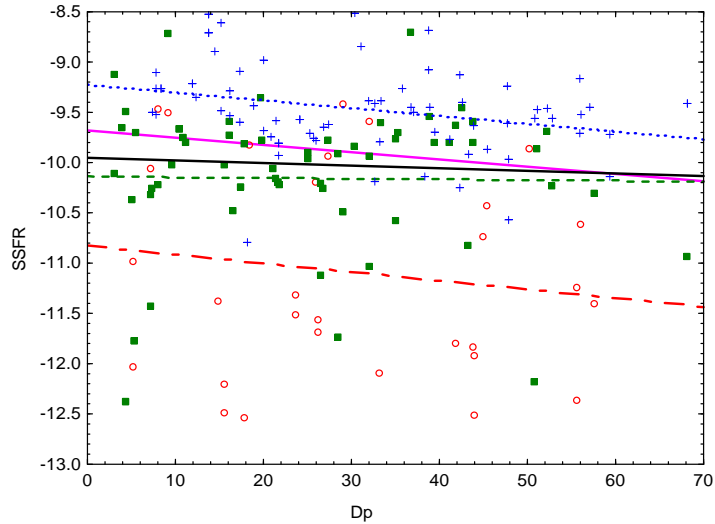}\\
\parbox{0.6\hsize}{\caption{SSFR vs. $D\rm{p}$ grouped by
morphologies. Points and lines marking
are the same as in Fig.~\ref{SSFRvsdVmorph}.}\label{SSFRvsDpmorph}}
\end{figure}

Figure~\ref{SSFRvsDpmorph} shows the dependence of SSFR on $D\rm{p}$ of a pair in AGN--removed sample.
If we do not categorize the sample by morphologies,
we see 0.7~dex increase of SSFR from larger $D\rm{p}$ to smaller ones.
In our sample $D\rm{p}$ is not biased by morphologies compared to ${\rm d}V$.
After grouping by morphologies we have about 0.5~dex difference of SSFRs
in closer pairs compared to the wider ones.
$D\rm{p}$ correlates stronger with the perturbation level $P$ than ${\rm d}V$.
Meanwhile, SSFRs correlate weakly with perturbation.
We interpret this partially as the result of two factors.
First, $P$ is biased by morphology: it is easier to detect disturbance of spiral galaxies,
and more difficult when galaxy is of early-type or irregular.
Second, peaks of SF usually coincide with pericenter passages
according to models \cite{dimatteo08}, however, the perturbations can be delayed.

\begin{figure}[t]
\centering
\includegraphics[width=0.6\hsize]{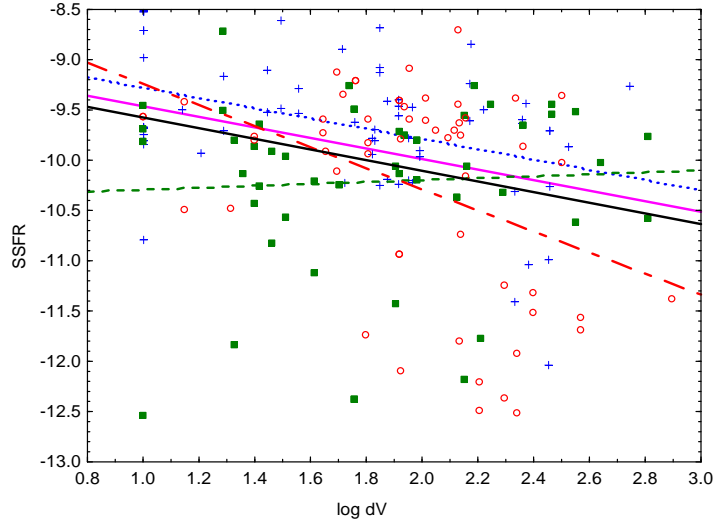}
\parbox{0.6\hsize}{\caption{SSFR vs. $log({\rm d}V)$ for
subsamples with large $\Sigma$ (red blank circles,
dashed-dotted line), medium $\Sigma$ (green filled
squares, dashed line), and low $\Sigma$
(blue crosses, dotted line). Two best-fit lines
for all galaxies (black bottom solid line) and
AGN--removed sample (purple upper solid
line) are drawn.}\label{SSFRvsdVSigma}}
\end{figure}
\begin{figure}[t]
\centering
\includegraphics[width=0.6\hsize]{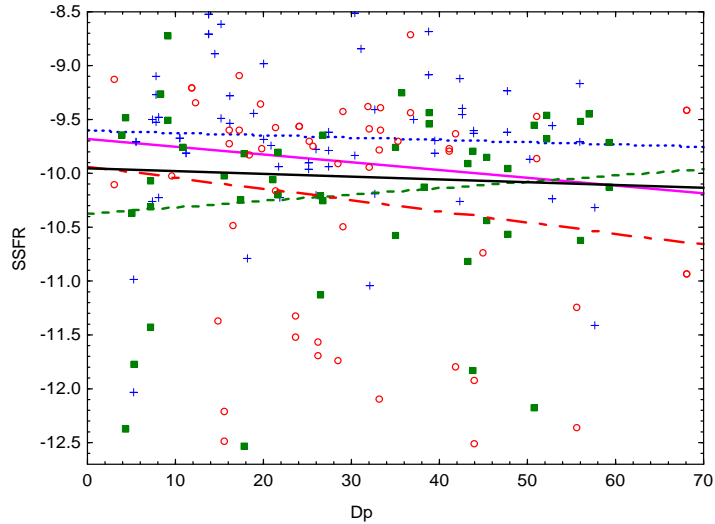}
\parbox{0.6\hsize}{\caption{SSFR vs. $D\rm{p}$ grouped by large-scale
density $\Sigma$. Points and lines
marking are the same as in Fig.~\ref{SSFRvsdVSigma}.}\label{SSFRvsDpSigma}}
\end{figure}

The dependence of SSFR on ${\rm d}V$ grouped according to
the density of large-scale environment $\Sigma$ is shown in Fig.~\ref{SSFRvsdVSigma}.
There is an increase of SSFR for all environments by about 0.5~dex.
We do not confirm the results in \cite{alonso06,ellison10}
which suggested stronger increase of SF in medium- and low-density environments than in high-density ones.
The dependence of SSFR on $D\rm{p}$ in environments with different $\Sigma$ is shown in Fig.~\ref{SSFRvsDpSigma}.
There is no trend of SSFR increase in medium-density environments,
while there is about 0.5~dex increase of SSFR in high-density environments
and weaker trend in low-density ones.
Binning both by morphologies and large-scale environments in larger samples
can make possible to separate the effects of morphologies
and large-scale environment densities on SSFR increase.
We tried also to group galaxies by their global colors
instead of morphologies to see variations of SSFR vs. ${\rm d}V$ and $D\rm{p}$
within each color group.
However, there is no variation.
This is probably because colors already depend on SSFR,
and variation of color reflects variation of SF even within each morphological class.

\begin{figure}[t]
\centering
\includegraphics[width=0.6\hsize]{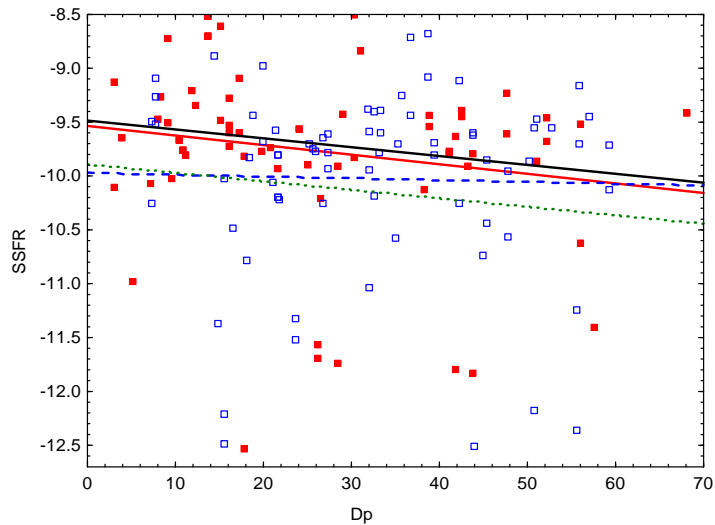}
\parbox{0.6\hsize}{\caption{SSFR vs. $D\rm{p}$ for major
interactions (red filled squares, solid line
best fit), minor interactions (blue blank
squares, dashed line), brightest
components of major interactions
(black upper solid line), and brightest
components of minor interactions
(green dotted line).}\label{SSFRvsDpmajmin}}
\end{figure}
\begin{figure}[t]
\centering
\includegraphics[width=0.6\hsize]{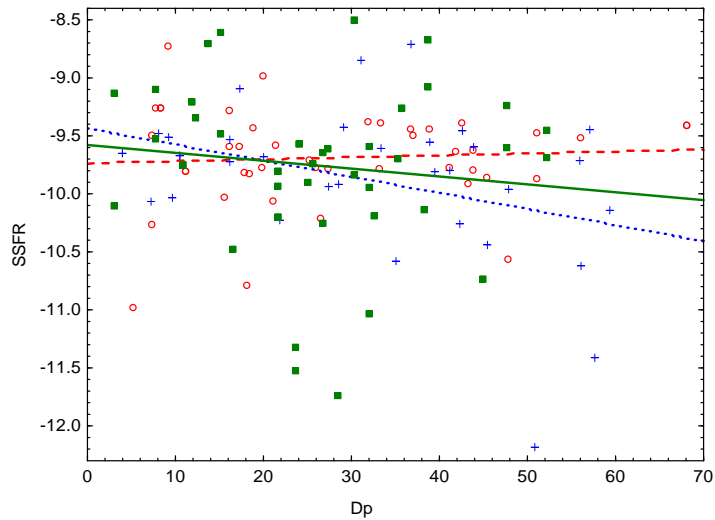}
\parbox{0.6\hsize}{\caption{SSFR vs. $D\rm{p}$ for galaxies with
relatively earlier type neighbors (red blank
circles, dashed line), same type
neighbors (green filled squares, solid line),
and relatively later type neighbors
(blue crosses, dotted line).}\label{SSFRvsnmorph}}
\end{figure}

We study the impact of the luminosity ratio of pair members on SSFR increase
by dividing all the pairs into two categories.
Those with $log(L_{\rm{bright}}/ L_{\rm{faint}})\leq 0.6$
we call major interactions, and the rest we call minor ones (see also \cite{lambas03,woods07,cox08}).
This separation is biased neither by redshift,
nor by morphology or large scale environment.
Figure~\ref{SSFRvsDpmajmin} shows the SSFR--$D\rm{p}$ relation
for major and minor interactions,
also the subsamples of the brightest members of pairs are showed separately.
The major interactions are more effective in triggering SF than minor ones,
there is a 0.5~dex increase of SF in major interactions,
while there is no trend among minor interactions.
Figure~\ref{SSFRvsDpmajmin} shows also that
the brightest members of major interactions have higher SSFRs than pairs in average,
therefore the brightest members obtain extra SSFR
(if consider not-normalized SFR, it would be even more).
This results are in agreement with previous both observational and modeling data
\cite{lambas03,woods07,cox08,scudder12b,lambas12}
suggesting that tidal forces draw gas into the central regions of galaxies,
and the merger mass ratio is an important parameter defining the effectiveness of the tidal forces.

The impact of morphology of neighbor galaxy on SSFR is shown in Fig.~\ref{SSFRvsnmorph}.
Existence of earlier-type neighbor does not increase of SSFR,
while the same-type and later-type neighbor increases SSFR of a galaxy.
The extra SSFR is maximal if the neighbor galaxy is of later morphological type,
in this case the SSFR increases by about 0.8~dex.
Previous papers \cite{hwang10,park09} also obtained similar result.
This result supports the scenario where close interaction with a later type neighbor
can not only trigger gas inflow in earlier type galaxy,
but also be an additional source of gas fuel.

\subsection{Nuclear types vs. pair properties}

We study the fraction of galaxies with different nuclear BPT types
by grouping the spiral galaxies in the sample according to the $D\rm{p}$ and $P$.
The fraction of AGN galaxies in less separated pairs is larger than that in more separated pairs.
The difference is especially obvious when considering the perturbation level $P$.
Fraction of AGNs changes from 8~\% for unperturbed pairs ($P=0$)
to 55~\% for mergers ($P=3$) respectively.
On the contrary, the fraction of star-forming galaxies changes from 69~\% to 45~\% in the same groups.
This result (see also \cite{ellison11a,alonso07}) indicates that
while both AGN and SF can be triggered by interactions,
AGNs ``prefer'' stronger interactions.
Different timings of starbursts and AGN events can explain this result \cite{ellison08,wild10,hopkins12}.

The fraction of AGNs in barred spirals is not significantly different from that in spirals without bars.
The fraction of AGNs in major interactions is about 4 times larger than
that in minor ones ($32 \pm 7~\%$ compared to $8 \pm 5~\%$), showing results similar to \cite{ellison11a}.

\subsection{Correlations between properties of pair members}

We study whether there is a correlation between SSFRs of galaxies in a pair.
First, we do this by grouping pairs according to
the environments with the same large-scale density $\Sigma$.
There are statistically significant correlations between SSFRs of pair members
within low- and high-density environments ($p = 0.00003$ and $p = 0.0003$ respectively)
with $r\sim0.6$ correlation coefficient.
Second, we also study correlation between SSFRs of galaxies in a pair,
by filtering the same-morphology pairs and grouping them according to morphologies.
The only statistically significant correlation we found is
between SSFRs of galaxies within S0/a--Sab ($r = 0.95$, $p=0.03$).
In all the other groups of galaxies there is no any correlation.
We explain the difference between these two results by the following reasons:
the statistics by grouped morphologies has smaller sample
than the statistics grouped by environments,
and the correlations between enhanced SSFRs of pair members due to interactions are generally weak.

We study BPT--BPT correlations between pair members.
For that purpose, we compare probability $p_{ij}$
($i$, $j$ = passive, LINER, starforming, composite, AGN)
with pure probability $p_{j}$,
where $p_{ij}$ is the conditional probability of a galaxy with BPT type $i$
to have a neighbor with BPT type $j$,
and $p_{j}$ is probability for a random galaxy to have BPT type $j$.
Table~\ref{BPTvsBPT} shows the coupling coefficients
$\gamma_{ij} = p_{ij} / p_{j}$.
We calculated the variances of the coefficients by random generation of
all possible pairs with mean numbers and standard deviations basing on existing numbers.
For each spectral type, there is a tendency to have an increased probability of a neighbor with the same BPT type.
Especially that is noticeable regarding passive galaxies and AGNs:
passive galaxies are about 6 times more likely to be found near another passive galaxy,
AGNs are 3 times more probable to be near another AGN.
While our results suffer from low number statistics,
we speculate that star-forming galaxies decrease the probability
to have a passive, LINER or AGN neighbor.
On the other hand, passive galaxies can increase the likelihood to have AGN or LINER neighbors.
It is not difficult to calculate, that sample selection criteria
does not decrease the diagonal coefficients,
so these values in the Table~\ref{BPTvsBPT} remain real for any sample
($\gamma$ coefficient of passive--passive pairs will be much more).
Previous results showed that interaction-triggered SF and AGNs often are
correlated between components in major interactions, e.g. \cite{liu12,scudder12b},
and have tendency to appear in the environments of galaxies with similar properties,
i.e. Seyfert galaxies tend to have the highest probability of having another Seyfert galaxy in the neighborhood,
HII galaxies tend to have the highest probability of having another HII galaxy in the neighborhood, etc., e.g. \cite{kollatschny12}.
We explain all the above mentioned correlations between BPT types as a result of two main factors.
First, morphologies of galaxies are correlated:
passive galaxies, AGNs and LINERs are more likely to be found among earlier type galaxies,
while star-forming galaxies are more likely to be later type galaxies \cite{ho08}.
In this respect, correlation of morphologies between pair members
automatically creates correlation of BPT types.
Second, stronger interactions increase the likelihood of pair members
to have nuclear activity of same types.

\begin{table}[t]
\begin{center}
\parbox{0.57\hsize}{\caption{The coupling coefficients $\gamma_{ij} = p_{ij} / p_{j}$ with their variances,
where $i$ corresponds to the target galaxy, $j$ to the neighbor.}\label{BPTvsBPT}}
\tabcolsep 1.3pt
\begin{tabular}{lccccc}
\hline
\hline
~&\multicolumn{5}{c}{Target}\\
\hline
Neighbor&PS&L&SF&C&AGN\\
\hline
\,\,\,\, PS&$~5.6\pm1.2~$&$~ 1.6\pm1.2~$& $~0.3\pm0.1~$&$~ 0.7\pm0.4~$& $~1.6\pm1.4~$\\
\,\,\,\, L&$~1.9\pm1.5~$&$~ 1.3\pm1.2~$& $~0.7\pm0.3~$&$~ 0.9\pm0.7~$& $~1.6\pm1.5~$\\
\,\,\,\, SF&$~0.2\pm0.1~$&$~ 0.8\pm0.3~$& $~1.1\pm0.1~$&$~ 0.8\pm0.2~$& $~0.6\pm0.4~$\\
\,\,\,\, C&$~0.6\pm0.5~$&$~ 0.8\pm0.7~$& $~0.9\pm0.2~$&$~ 1.5\pm0.5~$& $~1.0\pm1.0~$\\
\,\,\,\, AGN&$~1.4\pm1.1~$&$~ 1.5\pm1.6~$& $~0.5\pm0.3~$&$~ 1.7\pm1.0~$& $~3.3\pm3.2~$\\
\hline
\\
\end{tabular}
\end{center}
\end{table}

\section{Summary}
\label{summary}

We studied pairs containing MRK galaxies,
conducted both multivariate and one-dimensional statistics,
and came to the following main conclusions:

\begin{description}
\item[(\emph{i})]
Projected radial separation $D\rm{p}$ between galaxies correlates with the perturbation level $P$ of the pairs.
Both parameters do not correlate with the LoS velocity difference ${\rm d}V$ of pair members.
$D\rm{p}$ and $P$ are better measures of interaction strength than ${\rm d}V$.
The latter correlates with the density of large-scale environment and
with the morphologies of galaxies.
\item[(\emph{ii})]
Both galaxies in a pair are of the same nature, the only difference is that
MRK galaxies are usually brighter than their neighbors in average by 0.9~mag.
This result supports the existence of correlation between properties of paired galaxies.
\item[(\emph{iii})]
SSFRs of galaxies in pairs with smaller $D\rm{p}$ or ${\rm d}V$ is in average 0.5~dex higher
than that of galaxies in pairs with larger $D\rm{p}$ or ${\rm d}V$.
These trends are stronger when considering $D\rm{p}$, rather than ${\rm d}V$ and $P$.
These trends exist within all groups selected by morphologies and
within groups having low- and high-density environments.
\item[(\emph{iv})]
Major interactions/mergers trigger SF more effectively than minor ones.
The brightest components of pairs gain more SSFR than the fainter ones.
Closeness of a neighbor with same and later morphological type increases the SSFR,
while earlier-type neighbors do not increase SSFR.
\item[(\emph{v})]
The fraction of AGNs is higher in more perturbed pairs and pairs with smaller $D\rm{p}$.
AGNs typically are in stronger interacting systems than star-forming and passive galaxies.
Major interactions/mergers trigger AGNs more effectively than minor ones.
\item[(\emph{vi})]
The correlations between SSFRs of pair members within fixed environments have medium strength.
The correlations between SSFRs of pair members are weaker when considering same-morphology pairs.
Galaxy with given nuclear spectral type tend to increase
the probability of having a neighbor with similar nuclear properties.
We suspect that the presence of passive galaxy reduces the probability
to find a star-forming neighbor and increases the probability to find an AGN or LINER neighbor.
\end{description}

\begin{acknowledgement}
\,\ \\
\,\ \\
A.R.P. and A.A.H. acknowledge the hospitality of the Institut d'Astrophysique de Paris
(France) during their stay as visiting scientists supported by the Collaborative Bilateral Research
Project of the State Committee of Science (SCS) of the Republic of Armenia and the French
Centre National de la Recherch\'{e} Scientifique (CNRS).
This work was made possible in part by
a research grant from the Armenian National Science and Education Fund (ANSEF) based in
New York, USA.
This research made use of the NASA/IPAC Extragalactic Database, which is
available at \texttt{http://ned.ipac.caltech.edu}, and operated by the Jet Propulsion Laboratory,
California Institute of Technology, under contract with the National Aeronautics and Space
Administration. Funding for SDSS-III has been provided by the Alfred P. Sloan Foundation, the
Participating Institutions, the National Science Foundation, and the US Department of Energy
Office of Science. The SDSS-III web site is \texttt{http://www.sdss3.org}.
SDSS-III is managed by the
Astrophysical Research Consortium for the Participating Institutions of the SDSS-III
Collaboration including the University of Arizona, the Brazilian Participation Group,
Brookhaven National Laboratory, University of Cambridge, Carnegie Mellon University, University of Florida,
the French Participation Group, the German Participation Group, Harvard University,
the Instituto de Astrofisica de Canarias, the Michigan State/Notre Dame/JINA Participation Group,
Johns Hopkins University, Lawrence Berkeley National Laboratory,
Max Planck Institute for Astrophysics, Max Planck Institute for Extraterrestrial Physics,
New Mexico State University, New York University, Ohio State University, Pennsylvania State University,
University of Portsmouth, Princeton University, the Spanish Participation Group, University of Tokyo,
University of Utah, Vanderbilt University, University of Virginia, University of Washington, and Yale University.
\end{acknowledgement}

%

\end{document}